\documentclass[lettersize,journal]{IEEEtran}
\usepackage{amsmath,amsfonts}
\usepackage{algorithmic}
\usepackage{algorithm}
\usepackage{array}
\usepackage[caption=false,font=normalsize,labelfont=sf,textfont=sf]{subfig}
\usepackage{textcomp}
\usepackage{stfloats}
\usepackage{url}
\usepackage{verbatim}
\usepackage{graphicx}
\usepackage{cite}
\usepackage{makecell} 
\usepackage{amssymb} 
\newtheorem{definition}{Definition}
\newtheorem{theorem}{Theorem}
\usepackage{enumitem}
\usepackage{graphicx}    
\usepackage{subfig} 
\usepackage{hyperref}

\begin{document}

\title{Flexible Threshold Multi-client Functional Encryption for Inner Product in Federated Learning}

\author{

	Ruyuan Zhang, and Jinguang Han,~\IEEEmembership{Senior Member,~IEEE,} and Liqun Chen,~\IEEEmembership{Senior Member,~IEEE}
	\thanks{Ruyuan Zhang is with the School of Cyber Science and Engineering, Southeast University, Nanjing 211189, China (e-mail: ruyuanzhang@seu.edu.cn). }   
	\thanks{Jinguang Han is with the School of Cyber Science and Engineering, Southeast University, Nanjing 211189, China, 
	and also with the Wuxi Campus, Southeast University, Wuxi 214125, China (e-mail: jghan@seu.edu.cn).}
	\thanks{Liqun Chen is with the Department of Computer Science, University of Surrey, Guildford, Surrey GU27XH, United Kingdom (e-mail: liqun.chen@surrey.ac.uk).}
	
}

%



\maketitle

\begin{abstract}
	Federated learning (FL) is a distributed machine learning paradigm that enables multiple clients to collaboratively train a shared model without  disclosing their local data. 
	To address privacy issues of gradient, several privacy-preserving machine-learning schemes based on multi-client functional encryption (MCFE) have been proposed.
	However, existing MCFE-based schemes cannot support client dropout or flexible threshold selection, which are essential for practical FL. 	
	In this paper, we design a flexible threshold multi-client functional encryption for inner product (FTMCFE-IP) scheme, where multiple clients generate ciphertexts independently without any interaction.				
	In the encryption phase, clients are able to choose a threshold flexibly without reinitializing the system.
	The decryption can be performed correctly when the number of online clients satisfies the threshold.
	An authorized user are allowed to compute the inner product of the vectors associated with his/her functional key and the ciphertext, respectively, but cannot learning anything else.
	Especially,  the presented scheme supports clients drop out.
	Furthermore, we provide the definition and security model of our FTMCFE-IP scheme, and propose a concrete construction.	
	The security of the designed scheme is formally proven.
	Finally, we implement and evaluate our FTMCFE-IP scheme. 
\end{abstract}


\section{Introduction}
	\IEEEPARstart{F}{ederated} learning (FL) \cite{Kairouz2021} is a promising paradigm that enables multiple devices to collaboratively train a shared model without revealing their raw data to a centralized server. 
	However, FL is prone to serious privacy risks, such as gradient leakage, where sensitive information can be inferred from the exchanged gradients. 
	To address this issue, several privacy-preserving FL frameworks have been proposed: FL based on secure multi-party computation (FL-SMPC) \cite{Chen2024} \cite{Liu2024}, FL based on  homomorphic encryption (FL-HE) \cite{Gong2024} \cite{Yan2024} \cite{Zhang2020}, FL based on differential privacy (FL-DP) and  FL based on functional encryption (FL-FE) \cite{Chang2023} \cite{Feng2024} \cite{QianDMCFE2024} \cite{GuanSAMFL2024} \cite{Qian2022} \cite{YuLightweight2025}. 	
	\begin{itemize}[label=-]
		\item In FL-SMPC frameworks, training parameters are shared among participants through a secret sharing technology. 
		To decrypt, a centralized server collects information from a certain number of participants to reconstruct the secret, and then proceeds with the  model training, which introduces extra interactions and increases communication costs.	
		
		\item In FL-HE frameworks, a centralized server can conduct arithmetic operations directly on ciphertexts, thereby updating model parameters with encrypted gradients and outputting an encryption of the training outcome. However, the server requires further decryption for obtaining a final global model, which leads to high computation costs.
		
		\item In FL-DP frameworks, each participant locally perturbs its data by injecting random noise into the gradients, and only transmits  randomized gradients to the central server.
		Both the server and any other participants cannot recover the original sensitive information from the shared gradients.
		However, the introduction of random noise inevitably degrades the predictive accuracy of intermediate models.
		
		\item In FL-FE frameworks, a central server is granted a functional decryption key, and can directly perform computation on ciphertexts.
		The server obtains a function value of a encrypted message, but reveals no additional information about the underlying plaintext.
		Compared with existing approaches, FL-FE avoids intensive interactions in FL-SMPC, high computational overhead in FL-HE, and accuracy loss in FL-DP. 	
		
	\end{itemize}	
	
	Multi-client functional encryption (MCFE), which enables multiple clients to encrypt their private data individually, has been employed in FL to protect gradient privacy \cite{Chang2023} \cite{QianDMCFE2024} \cite{Qian2022} \cite{YuLightweight2025}.	
	In FL,  MCFE allows multiple clients to generate their ciphertexts locally.
	A central server collects ciphertexts from clients to aggregate their data and train models while disclosing nothing about intermediate gradient parameters.
	MCFE scheme is user-friendly in FL, where clients have limited computing resources while the central server is equipped with high computation power.
		
	Inner product (IP) is a fundamental operation in FL, which is employed to calculate a weighted average of local models for model updating.
	Function encryption for inner product (FE-IP) scheme is a specific instance of FE that allows the computation of inner products directly on encrypted data.
	To support encryption of multiple data sources, MCFE for inner product (MCFE-IP) was introduced \cite{Shi2023}.
	However, MCFE-IP schemes suffer from the drawback that all clients must remain online, which is not practical application.	
	In FL, the convergence of model relies on the collaborative training of multiple clients.
	When devices go offline, the number of clients participating in training decreases, which leads to a reduction in both the frequency of model updates and the volume of data used for each update, thereby increasing the number of training rounds required for the model to meet the expected performance metrics.	
	Hence, device offline/dropout is significant and must be supported.  
	Unfortunately, existing MCFE-IP schemes did not address this issue, and cannot satisfy the requirements of FL.
	
	In this paper, we design a flexible threshold multi-client functional encryption for inner product (FTMCFE-IP) scheme, where a decryptor can compute the inner product of vectors associated with his/her secret key and the ciphertext generated by participating clients when the decryptor gets enough functional keys.
	The threshold value can be flexibly selected without reinitializing the system.	
	In addition, our FTMCFE-IP scheme supports clients drop out, which is more suitable for FL application.	

\subsection{Related Work}

	\begin{table*}[!t]
		\centering
		\caption{Feature Comparison}
		\label{Feature Comparison}
		\begin{tabular}{|c|c|c|c|c|c|c|}
			\hline
			Scheme& { Multi-input / Multi-client}& \makecell{Threshold  Setting}& {Non-interactive  Setup Phase}& {Client  Dropout}  \\ 
			\hline
			
			\cite{QianDMCFE2024}&  Multi-client& Static client threshold& $\times$& \checkmark \\		
			
			\cite{LiRobust2023}&  Multi-client& Static client threshold& $\times$& \checkmark  \\ 		
			
			\cite{Feng2024}& Multi-input& $\times$& \checkmark & $\times$ \\ 		
			
			\cite{GuanSAMFL2024}&  Multi-input& $\times$& \checkmark & $\times$ \\ 
			
			\cite{Qian2022}&  Multi-client& $\times$& \checkmark & \checkmark \\
			
			\cite{XuTAPFed2024}& Multi-client& Static user threshold& \checkmark & $\times$ \\ 
			
			\cite{YuLightweight2025}& Multi-client& $\times$ & \checkmark & \checkmark \\
			
			\cite{Chang2023}& Multi-client& $\times$ & \checkmark &$\times$ \\
			
			Ours& Multi-client& Flexible client threshold& \checkmark& \checkmark \\ 
			\hline
		\end{tabular}
	\end{table*}	

\subsubsection{Functional Encryption}
	Waters \cite{Waters2008} first presented the concept of FE, which overcomes the ``all-or-nothing" limitation (that is, decryptors either obtain the whole plaintext or recover nothing at all) in public-key encryption schemes.
	Assuming that a FE scheme supports a function family $\mathcal{F}$.
	A trusted authority TA generates a functional key $SK_{f}$ for a specified function $f \in \mathcal{F}$. 
	When provided a ciphertext $CT_{x}$ and a function key $SK_{f}$, a  decryptor can learn the functional value $f(x)$ but nothing else about the plaintext $x$.
	Formal definitions and security models of FE are introduced  by O'Neill \cite{ONeill2010} and Boneh et al. \cite{Boneh2011}. 
	Existing FE schemes can be classed into two types: 
	(1) FE with flexible and general function \cite{FEBC2017}\cite{FEMPC};
	(2) FE with efficient and specific function \cite{SimpleFE}\cite{LeeDMCFE2023}\cite{QFE2023}.	
	FE enables selective computations on encrypted data, and has been used in various applications for performing secure computations on encrypted data while protecting data confidentiality, including IoT \cite{IoT2023}, cloud computing \cite{Han2024}, etc.
	
	Considering data may come from multiple sources, Goldwasser et al. \cite{Goldwasser2014} first present a multi-input functional encryption (MIFE) scheme, which allows multiple parties to generate their ciphertexts separately.
	Nevertheless, MIFE schemes suffer from ``mix-and-match" attacks since ciphertexts from multiple parties can be aggregated to perform decryption.
	To resolve the above issue,  multi-client functional encryption was introduced \cite{Shi2023}, where all ciphertexts are bound with a specific label.
	Hence, ciphertexts associated with the same label can be combined to recover the plaintext.	
	Nevertheless, a key requirement for MCFE schemes is that the decryptor is required to collect the ciphertexts from all clients (namely, all clients need to be online and working normally), which inevitably weakens system robustness.
	To address this issue, Li \cite{LiRobust2023} proposed a robust decentralized MCFE  (DMCFE) scheme, which divided clients into two types, positive clients and negative clients. 
	Positive clients stand for participants that can work normally and is able to generate ciphertexts, while negative clients are opposite. 
	When the size of the positive client set is not less than a given threshold, decryption can be performed correctly.
	Similarly, Qian et al. \cite{QianDMCFE2024} designed a threshold DMCFE scheme.
	In \cite{QianDMCFE2024}, to support client dropout, each client applies a secret-sharing technology to pre-split its secret, and then sends secret shares to online clients.	
	When the client drops out,  the server can recover the secret by collecting sufficient shares from the remaining online clients.		
	However, in above schemes, clients must interactively initialize the system, which leads to large communication costs.
	Moreover, the threshold is fixed in the setup phase, and cannot be changed without reinitializing the system, which is not suitable for practical FL environment.
	
\subsubsection{FE for Federated Learning}

	Feng et al. \cite{Feng2024} introduced a MIFE scheme with proxy re-encryption for privacy-preserving FL (PPFL), allowing a semi-trusted proxy to carry out secure aggregations while not learning any intermediate parameter.
	Guan et al. \cite{GuanSAMFL2024} designed a novel aggregation mechanism based a dual-decryption MIFE scheme, which can produce distinct results using different keys.	 
	To resist ``mix-and-match" attacks, Qian et al. \cite{Qian2022} proposed a PPFL framework based MCFE, which provides privacy guarantees for clients’ gradients.
	Xu et al. \cite{XuTAPFed2024} present a threshold MCFE scheme, where model training requires the number of aggregators should be not less than a specific threshold.	
	Yu et al. \cite{YuLightweight2025} designed a lightweight PPFL framework based a flexible DMCFE scheme, which is suitable for resource-constrained devices, and supports both clients dropout and participation.	
	To prevent the leakage of intermediate parameters during uploading local models, Chang et al.\cite{Chang2023} designed a dual-mode DMCFE scheme to develop a novel PPFL framework.		
	
	The key differences between the schemes \cite{Feng2024}, \cite{GuanSAMFL2024}, \cite{Qian2022}, \cite{XuTAPFed2024}, \cite{YuLightweight2025}, \cite{Chang2023}, \cite{QianDMCFE2024}, \cite{LiRobust2023} and our FTMCFE-IP scheme are summarized below: 
	(1) the designed scheme is able to resist ``mix-and-match" attacks, but the scheme \cite{Feng2024} \cite{GuanSAMFL2024} cannot solve it;
	(2) our scheme can support flexible client threshold, but the schemes \cite{QianDMCFE2024} \cite{LiRobust2023} \cite{XuTAPFed2024}  can only support static client/user threshold.
	The schemes \cite{Feng2024}, \cite{GuanSAMFL2024}, \cite{Qian2022}, \cite{XuTAPFed2024}, \cite{YuLightweight2025} and \cite{Chang2023} do not support threshold.	
	(3) the proposed scheme allows clients to drop out, but the schemes \cite{Feng2024} \cite{GuanSAMFL2024}, \cite{XuTAPFed2024}, \cite{Chang2023} cannot support it.
	
	We present a comparison of the features of our FTMCFE-IP scheme with related schemes in Table \ref{Feature Comparison}, focusing on multi-input/multi-client setting, threshold setting, non-interactive setup phase, and client dropout.

\subsubsection{Our Contributions}
	Our FTMCFE-IP scheme offers the following features: 
	(1) multiple clients co-exist and locally generate their ciphertexts; 
	(2) to resist ``mix-and-match attacks", all ciphertexts are bound to a specified label, ensuring preventing unauthorized recombination;
	(3) a key holder can calculate an inner product of the two vectors embedded respectively in his/her ciphertexts and decryption key when the number of online clients reaches the specified threshold;
	(4) threshold value can be flexibly specified without requiring system reinitialization;
	(5) no interaction is required at setup phase;
	(6) clients can drop out.		
	
	Our contributions are outlined as follows: 
	(1) we provide the definition of the FTMCFE-IP scheme, and also give its security model;
	(2) the concrete construction is built based on an asymmetric (Type-III) pairing;
	(3) the security proof is formally reduced to well-known complexity assumption;
	(4) we present an implementation of our FTMCFE-IP scheme, and provide its efficiency analysis.

\section{Organization}

	The remainder of this paper is structured as follows.
	Section \ref{Preliminaries} shows the preliminaries applied throughout this paper.
	Section \ref{Construction} details our FTMCFE-IP scheme.
	Section \ref{Security} discusses the security proof.
	Section \ref{Experimental} reports the experimental evaluation.
	Section \ref{Conclusion} summarizes the paper, and points out future research directions.

\section{Preliminaries} \label{Preliminaries}
	This section presents the preliminaries for our FTMCFE-IP scheme, and Table \ref{symbols} summarizes all symbols used in the paper.    
	
	\begin{table}[!t]
		\centering
		\caption{Symbols}
		\begin{tabular}{l | l}
			\hline
			Symbols & Explanations \\ \hline
			$1^{\kappa}$ & a security parameter \\ 
			$t$ & a threshold value \\
			$l$ & a label \\
			$n$ & the number of clients \\ 
			$x_i$ & a plaintext \\
			$\mathbf{a}$ & a vector \\
			$\mathbf{A}$ & a matrix \\ 
			$\mathbf{x}$ & a plaintext vector\\
			$\mathbf{y}$ & a function vector \\
			$sk_i$ & a secret key \\ 
			$ek_i$ & an encryption key \\
			$\mathcal{A}$ & an adversary \\
			$\mathcal{B}$ & a simulator \\
			$\mathcal{C}$ & a challenger \\
			$\mathcal{HS}$ & a set of honest clients \\
			$\mathcal{CS}$ & a set of corrupted clients \\					
			FL & federated learning \\
			$mpk$ & public parameters \\						
			$CT_{i,l,t}$ & a ciphertext corresponding to $x_i$ under $l$ and $t$ \\
			$SK_{i,\mathbf{y}}$ & a partial functional key related with $\mathbf{y}$  \\
			PPT & probabilistically polynomial time \\	
			MCFE &  multi-client function encryption\\
			FTMCFE-IP & \makecell[l]{flexible threshold  multi-client functional\\ encryption  for inner product} \\ \hline
		\end{tabular}
		\label{symbols}
	\end{table}
	
\subsection{Bilinear Groups}
	
	\begin{definition}[Bilinear Groups \cite{IBE}]
		$G,\hat{G}$ and $G_T$ denote three multiplicative cyclic groups with prime order $p$.
		$g, \hat{g}$ represent generators of $G,\hat{G}$, respectively.
		$e: G \times \hat{G} \rightarrow G_T$ is a bilinear map which satisfies the properties listed below:
		\begin{enumerate}[label=-]
			\item Bilinearity. If $g \in G$ and $\hat{g} \in \hat{G}$, the equation
			$e(g^s,\hat{g}^z)= e(g,\hat{g})^{sz} = e(g^z,\hat{g}^s)$ holds for any $z,s \in Z_p$.
			
			\item Non-generation. $e(g,\hat{g}) \neq 1$ for any $g \in G$ and $\hat{g} \in \hat{G}$.
			
			\item Computability. $e(g,\hat{g})$ can be calculated efficiently for any $g \in G$ and $\hat{g} \in \hat{G}$.
		\end{enumerate}			 
	\end{definition}
	
	$\mathcal{BG} (1^{\kappa}) \rightarrow \left( G,\hat{G},G_T,e,p \right)$ represents a bilinear group generator, which takes a security parameter $1^{\kappa}$ as input, and returns a bilinear group $\left( G,\hat{G},G_T,e,p \right)$ with prime order $p$ and map $e: G \times \hat{G} \rightarrow G_T$.  
	Specifically, 
	for a matrix $\mathbf{A} = \left( a_{i,j}\right) \in Z_p^{2 \times 2}$, we define 
	\begin{equation*}
		g^{\mathbf{A}} = 
		\begin{bmatrix}
			g^{a_{11}} & g^{a_{12}} \\
			g^{a_{21}} & g^{a_{22}}
		\end{bmatrix} \in G^{2 \times 2},
		\hat{g}^{\mathbf{A}} = 
		\begin{bmatrix}
			\hat{g}^{a_{11}} & \hat{g}^{a_{12}} \\
			\hat{g}^{a_{21}} & \hat{g}^{a_{22}}
		\end{bmatrix} \in \hat{G}^{2 \times 2}.
	\end{equation*}	
	For matrices $\mathbf{A} = \left( a_{i,j}\right) \in Z_p^{2 \times 2}$ and $\mathbf{B} = \left( b_{i,j}\right) \in Z_p^{2 \times 2}$, we define
	\begin{equation*}
		\begin{split}
			&e(g^{\mathbf{A}},\hat{g}^{\mathbf{B}}) = e \left( 
			\begin{bmatrix}
				g^{a_{11}} & g^{a_{12}} \\
				g^{a_{21}} & g^{a_{22}}
			\end{bmatrix},
			\begin{bmatrix}
				\hat{g}^{b_{11}} & \hat{g}^{b_{12}} \\
				\hat{g}^{b_{21}} & \hat{g}^{b_{22}}
			\end{bmatrix} 
			\right) \\	&= 
			\begin{bmatrix}
				e(g^{a_{11}},\hat{g}^{b_{11}}) e(g^{a_{12}},\hat{g}^{b_{21}}) & e(g^{a_{11}},\hat{g}^{b_{12}}) e(g^{a_{12}},\hat{g}^{b_{22}})\\
				e(g^{a_{21}},\hat{g}^{b_{11}}) e(g^{a_{22}},\hat{g}^{b_{21}}) & e(g^{a_{21}},\hat{g}^{b_{12}}) e(g^{a_{22}},\hat{g}^{b_{22}})
			\end{bmatrix} \\
			&= e(g,\hat{g})^{\mathbf{A} \mathbf{B}} \in G_T^{2 \times 2}. 
		\end{split}		
	\end{equation*}
	There exist three types of pairings:	
	Type I. $G = G$;		
	Type II. $G \neq \hat{G}$ with an efficient isomorphism map $\Phi: \hat{G} \rightarrow G$;		
	Type III. $G \neq \hat{G}$ without any efficient isomorphism map $\Phi: \hat{G} \rightarrow G$.	
	We will apply the Type III  pairing to build our FTMCFE-IP scheme due to its efficient performance.

\subsection{Complexity Assumptions}
	
	\begin{definition}[Decisional Diffie-Hellman Assumption (DDH) \cite{DDMCFE2020}]
		Let $G$ be a group with prime order $p$, and $g$ denotes a generator of $G$.
		Provided that $\left( g^a,g^r \right) \in G^{2}$, we say that DDH assumption holds on  $G$ if all probabilistic-polynomial time (PPT) adversaries $\mathcal{A}$ who is able to distinguish ${\left( g^{a},g^{r},g^{ar} \right)}$ and ${\left(  g^{a},g^{r},g^{z} \right)}$ with negligible advantage $\epsilon(\kappa)$, namely
		\begin{equation*}
			Adv_{\mathcal{A}}^{DDH} 
			= \left|
			\begin{array}{c}
				Pr[\mathcal{A}(g^{a}, g^{r}, g^{ar}) = 1] \\
				- Pr[\mathcal{A}(g^{a}, g^{r}, g^{z}) = 1]
			\end{array}
			\right|
			\leq \epsilon(\kappa),		
		\end{equation*}
		where $a,r,z \overset{R}{\leftarrow} Z_p.$
	\end{definition}
	
	Equivalently, the above assumption states that no adversary is able to distinguish, knowing $g^{a}$, a random value from the space of $g^{\mathbf{A}}$ for a matrix $\mathbf{A}=
	\begin{bmatrix}
		1 & {0} \\
		{0} & a
	\end{bmatrix}$, 
	from a random value in ${G}^{2 \times 2}$ with non-negligible advantage: $\left( g^{\mathbf{A}} \right)^{r}=g^{\mathbf{A}r} = 
	\begin{bmatrix}
		g^{r} & 0 \\
		0 & g^{ar}
	\end{bmatrix} 
	\approx 
	\begin{bmatrix}
		g^{r} & 0 \\
		0 & g^{z}
	\end{bmatrix}$.
	
	\begin{definition}[Multi-DDH \cite{DMCFE2018}] 
		Given that $\mathcal{R}_m = \{( D,( E_j,F_j = CDH(D,E_j) )_j )|D,E_j \xleftarrow{R} G, j\in [m]\}$ and $\mathcal{R}'_m = \{( D,( E_j,F_j ) )|D,E_j,F_j \xleftarrow{R} G, j\in [m]\}$.
		We say that the multi-DDH assumption holds on $G$ if all PPT adversaries running time $t$ can distinguish $\mathcal{R}_m$ and $\mathcal{R}'_m$ with advantage
		\begin{equation*}
			Adv_{\mathcal{A}}^{multi-DDH} \leq Adv_{\mathcal{A}}^{DDH} \left( t+4m\times t_{G} \right).			 
		\end{equation*}
	\end{definition}
	
	\begin{definition}[Symmetric eXternal Diffie-Hellman Assumption (SXDH) \cite{DDFE2025}]
		Let $\mathcal{BG} (1^{\kappa}) \rightarrow \left( G,\hat{G},G_T,e,p \right)$.		
		We say that the SXDH assumption holds on $(G,\hat{G},G_{T},e,p)$ if the DDH assumption hold on both the group $G$ and $\hat{G}$.
	\end{definition}

\subsection{System Model}
	
\begin{figure*}[!t] 
		\centering
		\includegraphics[height=10cm,width=14cm]{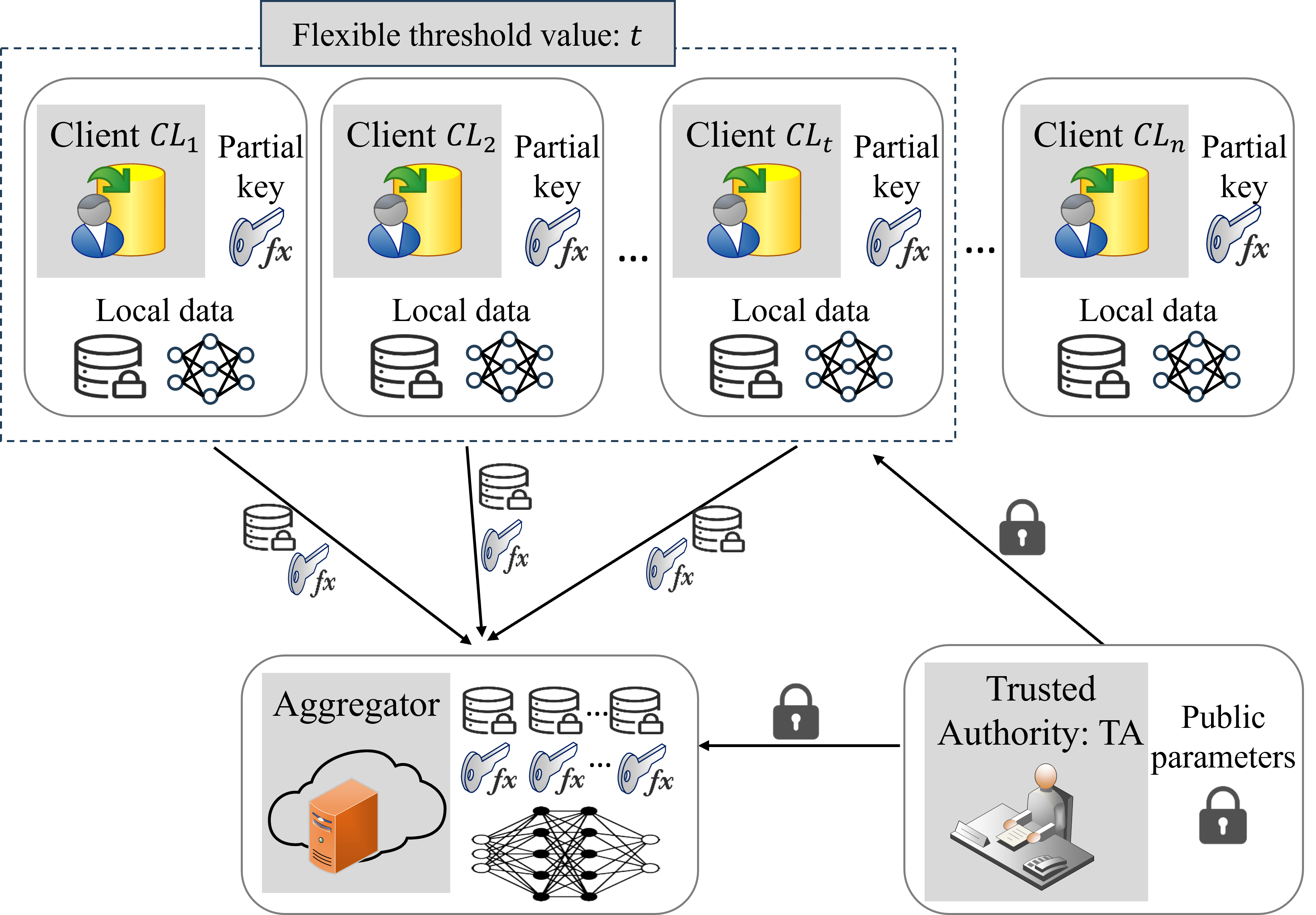}
		\caption{The framework of our FTMCFE-IP scheme.}
		\label{framework}
	\end{figure*}

	Fig \ref{framework} illustrates the framework of our FTMCFE-IP scheme.
	There exist three types of entities, including $n$ clients ($\{CL_i\}_{i\in[n]}$), a trusted authority (TA) and an aggregator. 	
	TA is responsible for generating public parameters.
	Each $\{CL_i\}_{i\in[n]}$ generates its secret key and encryption key, and calculates its ciphertext locally.
	The aggregator selects a set of clients, and requests partial function keys from these clients.
	Finally, the aggregator can output an value of inner product if and only if the number of partial functional keys it received reaches the specified threshold.
	
\subsection{Formal Definition}
	\begin{definition}[\textbf{Flexible Threshold  Multi-client Functional Encryption for Inner Product  (FTMCFE-IP)}]
		$\mathcal{F}=\{\mathcal{F}_\delta\}_{\delta \in \mathbb{N}}$ denotes a function family indexed by $\delta$ of functions $f: \mathcal{X}_{\delta,1} \times \ldots \times \mathcal{X}_{\delta,n_{\delta}} \rightarrow \mathcal{Y}_{\delta}$. Let $t \in [n]$ be a threshold and $\mathcal{L}=\{0,1\}^*$ a label set.	
		A FTMCFE-IP scheme for $\mathcal{F}$, $t$, $\mathcal{L}$ contains four algorithms: 
		
		$\mathbf{Setup}\left(1^\kappa,n\right)\rightarrow \left(mpk,\{sk_i,ek_i\}_{i\in [n]}\right)$:
		This algorithm is executed by TA and all online client $CL_i$.
		It takes a security parameter $1^\kappa$ and the number of client $n$ as input, and outputs public keys $mpk$ and secret/encryption keys $\{sk_i,ek_i\}_{i\in [n]}$.		
		The $mpk$ is implicitly used as input to other algorithms.
		
		$\mathbf{PKeyGen}(B,sk_i,\mathbf{y},t)\rightarrow SK_{i,\mathbf{y}}$:
		This algorithm is carried out by $CL_i$.
		It inputs a set of clients $B$, its secret key $sk_i$, a function vector $\mathbf{y}$ and a threshold $t$, and takes a partial functional key as output.		
		
		$\mathbf{Enc}(x_i,ek_i,t,l)\rightarrow CT_{i,l,t}$:
		This algorithm is performed by each client $CL_i$.
		It inputs a message $x_i$, a encryption key $ek_i$, a threshold $t$ and a label $l$, and outputs a ciphertext $ CT_{i,l,t}$.
		
		$\mathbf{Dec}(B,\mathbf{y},\{SK_{i,\mathbf{y}}\}_{i \in B},\{CT_{i,l,t}\}_{i \in B},l)\rightarrow \langle \mathbf{x},\mathbf{y} \rangle$:
		This algorithm is performed by an aggregator.
		It inputs an index set of clients $B$, a function vector $\mathbf{y}$, a set of partial functional keys $\{SK_{i,\mathbf{y}}\}_{i \in B}$, a group of ciphertexts $\{CT_{i,l,t}\}_{i \in B}$ and a label $l$, and outputs an inner product $\langle \mathbf{x},\mathbf{y} \rangle$.
		
		\noindent \textbf{Correctness.} A FTMCFE-IP scheme is correct, if 		
		\begin{equation*}
			\Pr \left[
			\begin{array}{l|l}
				\mathbf{Dec}(B,\mathbf{y},
				
				& \mathbf{Setup}\left(1^\kappa,n\right)\rightarrow (mpk, \\ 
				
				\{SK_{i,\mathbf{y}}\}_{i \in B}, &\{sk_i,ek_i\}_{i\in [n]}); \\
				
				\{CT_{i,l,t}\}_{i \in B},l)

				& \mathbf{PKeyGen}(B,sk_i,f,t) \\ 
				\rightarrow \langle \mathbf{x},\mathbf{y} \rangle; &\rightarrow SK_{i,\mathbf{y}};\\
				
				& \mathbf{Enc}(x_i,ek_i,t,l)\rightarrow CT_{i,l,t}; 
			\end{array}
			\right] = 1.
		\end{equation*}  
		
	\end{definition}
	
\subsection{Security Model}

	For every FTMCFE-IP scheme with $\mathcal{F}$, $\mathcal{L}$ and $t$, a security parameter $\kappa \in \mathbb{N}$, $b \in \{0,1\}$, an adversary $\mathcal{A}$, considering the experiments described in Fig.\ref{security model}.
	
	\begin{figure*}[!t] 
		\centering
		\fbox{			
			\begin{minipage}{\textwidth}
				\begin{flushleft} 
					\textbf{Experiment FTMCFE-IP-Expt}$^{b}(1^{\kappa})$:
				\end{flushleft}
				\begin{enumerate}[leftmargin=*,label={\textbullet}]
					\item \textbf{Setup.} 
					$\mathcal{A}$ chooses a set of corrupted clients $\mathcal{CS} \subset [n]$, and sends
					$\mathcal{CS}$ to challenger $\mathcal{C}$. 
					$\mathcal{C}$ executes $Setup\left(1^\kappa,n\right)\rightarrow \left(mpk,\{sk_i,ek_i\}_{i\in [n]}\right)$, and responses $mpk$ and $\{sk_i,ek_i\}_{i \in \mathcal{CS}}$ to $\mathcal{A}$.
					
					\item \textbf{Query.} $\mathcal{A}$ has adaptive access to the following types of oracles.					
					\begin{enumerate}[label=-]
						\item \textbf{QPKeyGen.} 
						$\mathcal{A}$ makes decryption key query with the tuple $\{\left(i,y_i\right)\}_{i\in [B]}$.
						$\mathcal{C}$ executes $PKeyGen(B,sk_i,\mathbf{y},t)\rightarrow SK_{i,\mathbf{y}}$, and responses $\{SK_{i,\mathbf{y}}\}_{i \in B}$, where $|B|=t>|\mathcal{CS}|$. 
						
						\item \textbf{QEncrypt.}  $\mathcal{A}$ makes encryption query with the tuples $\left(i,l,t,x_i^{0},x_i^{1}\right)$.
						$\mathcal{C}$ executes $Enc(x_i^{b},ek_i,t,l)\rightarrow CT_{i,l,t}^{b}$, and returns $CT_{i,l,t}^{b}$ to $\mathcal{A}$. 
						Any further query  $\left(i,l,t,*,*\right)$ will be ignored.					
					\end{enumerate}	
					\item \textbf{Finalize.}
					$\mathcal{A}$ outputs a guess $b'$ on $b$, and  this procedure returns a
					result $\alpha$ of the security game. 				
				\end{enumerate}
			\end{minipage}		
		}
		\caption{Security model of our FTMCFE-IP scheme.}
		\label{security model}
	\end{figure*}
	
	We set the output $\alpha=b'$ iff all the following conditions hold, where $\mathcal{HS} = [n]\setminus \mathcal{CS}$. 
	Otherwise, we set $\alpha$ as a random bit, i.e., $\alpha \in \{0,1\}$:		
	\begin{enumerate}
		\item for each $i \in \mathcal{CS}$, the \textbf{QEncrypt} queries with $\left(i,l,t,x_i^{0},x_i^{1}\right)$ satisfy $x_i^{0}=x_i^{1}$.
		
		\item for each $l \in \mathcal{L}$, either  $\mathcal{A}$ has made as least one \textbf{QEncrypt} query with $\left(i,l,t,x_i^{0},x_i^{1}\right)$ for each $i \in B$, or $\mathcal{A}$ has made no \textbf{QEncrypt} query with $\left(i,l,t,x_i^{0},x_i^{1}\right)$ for each $i \in B$.
		
		\item for each $l \in \mathcal{L}$, any \textbf{QEncrypt} query with $\{ \left(i,l,t,x_i^{0},x_i^{1}\right) \}_{i \in B}$ and each \textbf{QPKeyGen} query with $\{\left(i,y_i\right)\}_{i\in [B]}$, 
		we require $\sum_{i \in B}\langle x_i^{0},y_i \rangle = \sum_{i \in B}\langle x_i^{1},y_i\rangle$.			
	\end{enumerate}		
	\begin{definition}
		A FTMCFE-IP scheme is static-IND-secure if for any $\mathcal{A}$ who is able to win the game describe in Fig \ref{security model} with negligible advantage $\epsilon(\kappa)$, namely		
		\begin{equation*}
			Adv^{\text{sta-IND}}(\mathcal{A})
			= \left|
			\begin{array}{c}
				\Pr[\alpha=1 \mid b=1] \\
				- \Pr[\alpha=1 \mid b=0]
			\end{array}
			\right|
			\leq \epsilon(1^\kappa).
		\end{equation*}		
	\end{definition}

\section{Concrete Construction of Our FTMCFE-IP scheme} \label{Construction}
\subsection{Degree-check Technique}

	The degree-check technique can be applied to verify the degree of a polynomial introduced by Garg et al. \cite{Silent2024}.
	Assuming a vector $\mathbf{b} = (b_1,b_2,...,b_n)$ contains at least $t$ non-zero coordinates (i.e., $||\mathbf{b}||_0 = |\{i|b_i \neq 0\}| \geq t$).
	Let $\mathbf{b}$ consist of the evaluations of a polynomial $B(x)$.
	It has at most $n-t$ zeros (i.e., $|\{x|B(x) = 0\}| \leq t$) if the polynomial $B(x)$ has at least $t$ non-zeros (i.e., $|\{x|B(x) \neq 0\}| \geq t$).
	Hence, the degree $B(x)$ is at most $n-t$ (i.e., $deg(B(x)) \leq n-t$).
	Given $\{g_1^{\gamma^{j}}\}_{j\in[n]}, \{g_2^{\gamma^{j}}\}_{j\in[n]}, \{\hat{g_1}^{\gamma^{j}}\}_{j\in[n]}$, requesting a committed polynomial $\overline{B(x)} = B(x) \cdot x^t$ from a prover, the verifier can check if $deg(B(x)) \leq n-t$ by checking if
	\begin{equation*}
		e(g_2^{\overline{B(\gamma)}}, \hat{g_1})  \stackrel{?}{=}
		e(g_2^{\gamma^t},\hat{g_1}^{B(\gamma)}).
	\end{equation*}
	In our scheme, all clients are organized into a vector $\mathbf{b} = (b_1,b_2,...,b_n)$  according to their index.
	Set $b_i = 1$ if the client $CL_i$ is online or the client $CL_i$ is able to participate in, otherwise set $b_i = 0$.
	We will check if the number of online clients has at least $t$ by running a degree check on $B(x)$.
	
\subsection{Main Construction}

	In this section, we provide the concrete construction of our FTMCFE-IP scheme, which contains four algorithms, namely
	$\mathbf{Setup}$, $\mathbf{PKeyGen}$, $\mathbf{Enc}$, and $\mathbf{Dec}$ algorithms, as shown in Fig \ref{setup} to Fig \ref{dec}. 
		\begin{figure*}[!t]
		\centering
		\fbox{			
			\begin{minipage}{\textwidth}				
				{
					
					$\mathbf{Setup}\left(1^\kappa,n\right)\rightarrow \left(mpk,\{sk_i,ek_i\}_{i\in [n]}\right)$:	
					Suppose there exist $n$ clients $\{CL_1,CL_2,...,CL_n\}$.
					Let $\mathcal{BG}\rightarrow \left( G,\hat{G},G_T,e,p \right)$.
					$g_0,g_1,g_2$ denote generators of $G$, and $\hat{g_0},\hat{g_1},\hat{g_2}$ stand for generators of $\hat{G}$. 
					$H_1: Z_p^*
					\rightarrow \hat{G}^{2 \times 2}$ and $H_2:\{0,1\}^{*} \rightarrow G^{2 \times 2}$  are two cryptographic hash function.
					Each client $CL_i$ picks $w_i \in Z_p$, 
					$\mathbf{T}_i = \begin{bmatrix} t_i^{1} & 0 \\ 0 & t_i^{2} \end{bmatrix} \in {Z}_p^{2 \times 2}$
					and $\mathbf{S}_i = \begin{bmatrix} s_i^{1} & 0  \\ 0 & s_i^{2} \end{bmatrix} \in {Z}_p^{2 \times 2}$ randomly.	
					It sets its secret key as $sk_i=\left(\mathbf{S}_i,\mathbf{T}_i,w_i \right)$, and encryption keys $ek_i=\left(\mathbf{S}_i,\mathbf{T}_i, w_i \right)$.	
					TA samples $\gamma \in Z_p^*$, and generates $g_1^{\gamma},g_1^{\gamma^{2}},...,g_1^{\gamma^{n}},g_2^{\gamma},g_2^{\gamma^{2}},...,g_2^{\gamma^{n}}, \hat{g}_1^{\gamma},\hat{g}_1^{\gamma^{2}},...,\hat{g}_1^{\gamma^{n}}$. 
					Hence, public parameters are 
					\begin{equation*}
						mpk = \left(H_1,H_2, G,\hat{G},G_T,e,p,g_0,g_1,\hat{g_0},\hat{g_1},\hat{g_2},\{g_1^{\gamma^{j}}\}_{j\in[n]}, \{g_2^{\gamma^{j}}\}_{j\in[n]}, \{\hat{g_1}^{\gamma^{j}}\}_{j\in[n]} \right).
					\end{equation*}			
				}				
			\end{minipage}			
		}		
		\caption{The $\mathbf{Setup}$ algorithm of our FTMCFE-IP scheme}
		\label{setup}
	\end{figure*}	
	\begin{figure*}[!t]
		\centering
		\fbox{			
			\begin{minipage}{\textwidth}				
				{
					$\mathbf{PKeyGen}(B,sk_i,\mathbf{y},t)\rightarrow SK_{i,\mathbf{y}}:$
					The aggregator requests partial functional keys from a set of clients $\{CL_i\}_{i \in B}$, and broadcasts $\left(B,\mathbf{y}\right)$ to each $CL_i$, where $\mathbf{y} \in Z_p^{n}$ and $B \subseteq [n]$.	
					Each $CL_i$ generates partial functional key for an inner-product function defined by $\mathbf{y}$ as $f_{\mathbf{y}}(\mathbf{x})=\langle \mathbf{x},\mathbf{y}\rangle$.	
					Set a vector $\mathbf{b} = \left(b_1,...,b_n\right) \in \{0,1\}^n$, where $b_i = 1$ if $i \in B$ and $b_i = 0$ if $i \notin B$. 
					$\mathbf{b}$ can be treated as the evaluation form of the polynomial $B(x)$.
					Each $CL_i$ calculates the polynomial $B(x) = \prod_{i \notin [B]}(x-b_i) = a_{n-t}x^{n-t}+...+a_1x+a_0$, and commits it as $g_1^{B(\gamma)} = \prod_{i \in [n-t]} (g_1^{\gamma^i})^{a_i}$ and $\hat{g_1}^{B(\gamma)} = \prod_{i \in [n-t]} (\hat{g_1}^{\gamma^i})^{a_i}$.
					Then, it computes the polynomial $\overline{B(x)} = B(x)\cdot x^{t}$, and obtain $g_2^{\overline{B(\gamma)}}$.
					$CL_i$ uses its secret keys to calculate $\hat{g}^{\mathbf{H}_{\mathbf{y}}}=H_1(\mathbf{y})=\begin{bmatrix} \hat{g}^{h_{\mathbf{y}}^{1}} & 0 \\ 0 & \hat{g}^{h_{\mathbf{y}}^{2}} \end{bmatrix}\in \hat{G}^{2 \times 2}$ 
					and partial functional keys $SK_{i,\mathbf{y}} = \left( sk_{i,\mathbf{y}}^{(1)}, sk_{i,\mathbf{y}}^{(2)}, sk_{i,\mathbf{y}}^{(3)}, sk_{i,\mathbf{y}}^{(4)}, sk_{i,\mathbf{y}}^{(5)} \right)$ as follows \footnote{Only the hash value $\hat{g}^{\mathbf{H}_{\mathbf{y}}}$ is used, where $\mathbf{H}_{\mathbf{y}}$ is unknown.}.
					\begin{equation*}
						SK_{i,\mathbf{y}}
						=\left( \hat{g_0}^{\mathbf{S}_i\cdot y_i}\cdot\hat{g}^{\mathbf{T}_i \mathbf{H}_\mathbf{y}},
						\left( \hat{g}^{\mathbf{H}_{\mathbf{y}}} \cdot \hat{g_1}^{B(\gamma)} \right)^{\mathbf{T}_i},
						g_1^{B(\gamma)\mathbf{T}_i},
						\hat{g}^{\mathbf{H}_{\mathbf{y}}w_i},
						g_2^{\overline{B(\gamma)}\mathbf{T}_i}
						\right).
					\end{equation*}	
				}				
			\end{minipage}			
		}		
		\caption{The $\mathbf{PKeyGen}$ algorithm of our FTMCFE-IP scheme}
		\label{keygen}
	\end{figure*}	
	\begin{figure*}[!t]
		\centering
		\fbox{			
			\begin{minipage}{\textwidth}				
				{
					$\mathbf{Enc}(x_i,ek_i,t,l)\rightarrow CT_{i,l,t}:$
					The label $l$ is typically a time-stamp that uniquely identifies an encryption session.
					Prior to encryption, the participating clients jointly determine a threshold $t$.					
					To encrypt the message $x_i \in Z_p^*$ under $l$ and $t$, each $CL_i$ selects a threshold $t$, and calculates $g^{\mathbf{H}_l}=H_2(l)=\begin{bmatrix} g^{h_l^{1}} & 0 \\ 0 & g^{h_l^{2}} \end{bmatrix}\in G^{2 \times 2}$, 
					and generates the ciphertexts $CT_{i,l,t} =\left( C_{i,l,t}^{(1)},C_{i,l,t}^{(2)},C_{i,l,t}^{(3)},C_{i,l,t}^{(4)},C_{i,l,t}^{(5)} \right)$ as follows \footnote{Only the hash value $g^{\mathbf{H}_l}$ is used, where $\mathbf{H}_l$ is unknown.}.
					\begin{equation*}
						CT_{i,l,t}=\left( 
						\left(g^{\mathbf{H}_l}\right)^{\mathbf{S}_i} \cdot g_0^{x_i}, 
						\left( g^{\mathbf{H}_l}\cdot g_1^{w_i} \right)^{\mathbf{T}_i}, 
						\left( g_1\cdot g_2^{\gamma^{t}} \right)^{w_i}, 
						\hat{g_1}^{w_i},
						g_2^{\gamma^{t}\cdot \mathbf{T}_i}
						\right).
					\end{equation*}		
				}				
			\end{minipage}			
		}		
		\caption{The $\mathbf{Enc}$ algorithm of our FTMCFE-IP scheme}
		\label{enc}
	\end{figure*}	
	\begin{figure*}[!t]
		\centering
		\fbox{			
			\begin{minipage}{\textwidth}				
				{
					$\mathbf{Dec}(B,\mathbf{y},\{SK_{i,\mathbf{y}}\}_{i \in B},\{CT_{i,l,t}\}_{i \in B},l)\rightarrow \langle \mathbf{x},\mathbf{y} \rangle:$ 
					Received $\{SK_{i,\mathbf{y}}\}_{i \in B}$ and $\{CT_{i,l,t}\}_{i \in B}$, aggregator executes as follows.
					\begin{itemize}
						\item If $|B|<t$, it aborts.
						\item Otherwise, it calculates						
						\begin{equation*}
							\begin{split}					
								\prod_{i=1}^{B}\frac{
									e\left( C_{i,l,t}^{(1)},\hat{g_0}^{y_i} \right) \cdot 
									e\left( C_{i,l,t}^{(2)},\hat{g}^{\mathbf{H_y}}  \right) \cdot 
									e\left(sk_{i,\mathbf{y}}^{(3)},C_{i,l,t}^{(4)} \right) \cdot
									e\left(C_{i,l,t}^{(5)},sk_{i,\mathbf{y}}^{(4)} \right) \cdot
									e\left(sk_{i,\mathbf{y}}^{(5)},C_{i,l,t}^{(4)} \right)
								}
								{e\left( g^{\mathbf{H}_{l}},sk_{i,\mathbf{y}}^{(1)} \right) \cdot
									e\left( C_{i,l,t}^{(3) },sk_{i,\mathbf{y}}^{(2)} \right)
								} 
								= e\left( g_0,\hat{g_0} \right)^{\langle \mathbf{x},\mathbf{y} \rangle}.
							\end{split}			 
						\end{equation*}						
					\end{itemize}	
				}				
			\end{minipage}			
		}		
		\caption{The $\mathbf{Dec}$ algorithm of our FTMCFE-IP scheme}
		\label{dec}
	\end{figure*}
		
	\textit{Correctness.}		
	\begin{equation*}
		\begin{split}
			\mathcal{K}^{1} &= 
			\prod_{i=1}^{B} e\left(  C_{i,l,t}^{(1)},\hat{g_0}^{y_i} \right)
			= \prod_{i=1}^{B} e\left( \left(g^{\mathbf{H}_l}\right)^{\mathbf{S}_i} \cdot g_0^{x_i},\hat{g_0}^{y_i} \right) \\
			&= e\left( g,\hat{g_0} \right)^{\sum_{i \in B} \mathbf{H}_l \mathbf{S}_i y_i} \cdot e\left( g_0,\hat{g_0} \right)^{\langle \mathbf{x},\mathbf{y} \rangle},
		\end{split}			
	\end{equation*}		
	\begin{equation*}
		\begin{split}
			\mathcal{K}^{2} &= 
			\prod_{i=1}^{B} e\left( C_{i,l,t}^{(2)},\hat{g}^{\mathbf{H_y}}  \right) 
			= \prod_{i=1}^{B} e\left( \left( g^{\mathbf{H}_l}\cdot g_1^{w_i} \right)^{\mathbf{T}_i},\hat{g}^{\mathbf{H_y}} \right) \\
			&= e\left( g,\hat{g} \right)^{\sum_{i \in B} \mathbf{H}_l \mathbf{T}_i {\mathbf{H_y}}} 
			\cdot e\left( g_1,\hat{g} \right)^{\sum_{i \in B}w_i \mathbf{T}_i {\mathbf{H_y}}}, 
		\end{split}
	\end{equation*}	
	\begin{equation*}
		\begin{split}
			\mathcal{K}^{3} &=
			\prod_{i=1}^{B} e\left(sk_{i,\mathbf{y}}^{(3)},C_{i,l,t}^{(4)} \right)  
			= \prod_{i=1}^{B} e\left( g_1^{B(\gamma)\mathbf{T}_i}, \hat{g_1}^{w_i}  \right) \\		
			&= e \left( g_1,\hat{g_1} \right)^{\sum_{i \in B} B(\gamma)\mathbf{T}_i w_i },
		\end{split}
	\end{equation*}	
	\begin{equation*}
		\begin{split}
			\mathcal{K}^{4} &= 
			\prod_{i=1}^{B} e\left(C_{i,l,t}^{(5)},sk_{i,\mathbf{y}}^{(4)} \right) 
			= \prod_{i=1}^{B} e \left(g_2^{\gamma^{t}\cdot \mathbf{T}_i},\hat{g}^{\mathbf{H_y}w_i} \right) \\			
			&= e \left( g_2,\hat{g} \right)^{ \sum_{i \in B} \gamma^{t} \mathbf{T}_i \mathbf{H_y}w_i }, 				
		\end{split}			
	\end{equation*}	
	\begin{equation*}
		\begin{split}
			\mathcal{K}^{5} &=
			\prod_{i=1}^{B} e\left(sk_{i,\mathbf{y}}^{(5)},C_{i,l,t}^{(4)} \right) 				
			= \prod_{i=1}^{B} e\left( g_2^{\overline{B(\gamma)}\mathbf{T}_i}, 	\hat{g_1}^{w_i} \right) \\
			&= e\left( g_2,\hat{g_1} \right)^{\sum_{i \in B} \overline{B(\gamma)}\mathbf{T}_i w_i}, 
		\end{split}
	\end{equation*}	
	\begin{equation*}
		\begin{split}
			\mathcal{K}^{6} &=
			\prod_{i=1}^{B} e\left( g^{\mathbf{H}_{l}},sk_{i,\mathbf{y}}^{(1)} \right)
			= \prod_{i=1}^{B} e \left( g^{\mathbf{H}_{l}}, \hat{g_0}^{\mathbf{S}_i\cdot y_i}\cdot\hat{g}^{ \mathbf{T}_i \mathbf{h}_\mathbf{y}} \right) \\
			&= e \left( g, \hat{g_0} \right)^{\sum_{i \in B} \mathbf{H}_{l} \mathbf{S}_i y_i} \cdot e \left( g,\hat{g} \right)^{\sum_{i \in B} \mathbf{H}_{l} \mathbf{T}_i \mathbf{H_y} },
		\end{split}
	\end{equation*}
	and 
	\begin{equation*}
		\begin{split}
			\mathcal{K}^{7} &=
			 \prod_{i=1}^{B} e( C_{i,l,t}^{(3)},sk_{i,\mathbf{y}}^{(2)} ) 
			= \prod_{i=1}^{B} e ( ( g_1\cdot g_2^{\gamma^{t}} )^{w_i},( \hat{g}^{\mathbf{H}_{\mathbf{y}}} \cdot \hat{g_1}^{B(\gamma)} )^{\mathbf{T}_i} )  \\
			& =  e \left(g_1,\hat{g} \right)^{\sum_{i \in B} w_i \mathbf{H}_{\mathbf{y}} \mathbf{T}_i} \cdot 			
			e \left( g_1,\hat{g_1} \right)^{\sum_{i \in B} w_i B(\gamma) \mathbf{T}_i} \\
			& \quad \quad \cdot e \left( g_2,\hat{g} \right)^{\sum_{i \in B} \gamma^{t} w_i \mathbf{H}_{\mathbf{y}}\mathbf{T}_i} \cdot
			e \left( g_2, \hat{g_1} \right)^{\sum_{i \in B} \gamma^{t} w_i B(\gamma) \mathbf{T}_i}. 		
		\end{split}
	\end{equation*}
	Therefore, 	
	\begin{equation*}			
		\begin{split}
			 \quad 
			\prod_{i=1}^{B}\frac{
				\mathcal{K}^{1} \cdot  
				\mathcal{K}^{2} \cdot
				\mathcal{K}^{3} \cdot
				\mathcal{K}^{4} \cdot
				\mathcal{K}^{5} 
			}
			{\mathcal{K}^{6} \cdot
			 \mathcal{K}^{7}
			} 
			= e\left( g_0,\hat{g_0} \right)^{\langle \mathbf{x},\mathbf{y} \rangle}.
		\end{split}			 
	\end{equation*}	
\section{Security Analysis} \label{Security}
	We provide the security proof of our FTMCFE-IP scheme in this section.
	
	\begin{theorem}[sta-IND-Security]
		The above FTMCFE-IP scheme is sta-IND secure in the random oracle model if the SXDH assumption holds on $\left( G,\hat{G},G_T,e,p \right)$. 
		For any PPT adversary $\mathcal{A}$ with running time $t$, we have 
		\begin{equation*}
			\begin{split}
				{Adv}^{IND} (\mathcal{A}) & \leq 2Q_1 \cdot {Adv}^{\textit{DDH}}_{{G}}(t) + 2Q_2 \cdot {Adv}^{\textit{DDH}}_{\hat{G}}(t) \\ 
				&+ \frac{2Q_1 + 2Q_2}{p} + {Adv}^{\textit{DDH}}_{{G}}(t + 4Q_1 \times t_{{G}}) \\ 
				&+ 2 \cdot {Adv}^{\textit{DDH}}_{\hat{G}}(t + 4Q_2 \times t_{\hat{G}}),			
			\end{split}		
		\end{equation*}
		where $Q_1 \text{ and } Q_2$ denote  the number of  queries to hash oracles $H_1$ and $H_2$, respectively. 
		$t_{G}$ and $t_{\hat{G}}$ represent the computational time for performing an exponentiation in $G$ and $\hat{G}$, respectively.
	\end{theorem}
		
	\begin{IEEEproof}	
		
		\noindent
		\hangafter=1
		\setlength{\hangindent}{2em} 
		\textbf{Game} $\mathcal{G}_0$: 
		this game corresponds to the real game, where 	 $\hat{g}^{\mathbf{H}_{\mathbf{y}}}=H_1(\mathbf{y}) = \begin{bmatrix} \hat{g}^{h_{\mathbf{y}}^{1}} & 0 \\ 0 & \hat{g}^{h_{\mathbf{y}}^{2}} \end{bmatrix} \in \hat{G}^{2 \times 2}$ and $g^{\mathbf{H}_l}=H_2(l)=  \begin{bmatrix} g^{h_l^{1}} & 0 \\ 0 &  g^{h_l^{2}} \end{bmatrix} \in G^{2 \times 2}$ . 
		In addition, a set of corrupted clients $\mathcal{CS}$ is selected.
		
		\noindent
		\hangafter=1
		\setlength{\hangindent}{2em}
		\textbf{Game} $\mathcal{G}_1$: this game is same with the $\mathcal{G}_0$, with the exception that we simulate $H_1$ as a random oracle $RO_1$ which outputs a truly random matrix in $\hat{G}^{2 \times 2}$. 
		Set $H_1(\mathbf{y})=F'(\mathbf{y}) \in \hat{G}^{2 \times 2}$, where $F'$ is a random function onto $\hat{G}^{2 \times 2}$.
		We have
		\begin{equation*}
			Adv_0 = Adv_1.
		\end{equation*}
		
		\noindent
		\hangafter=1
		\setlength{\hangindent}{2em}
		\textbf{Game} $\mathcal{G}_2$: this game is same with the $\mathcal{G}_1$, with the exception that we simulate any new output of $RO_1$ as a truly random matrix in the space $\hat{g}^{\mathbf{A}}$ for $\mathbf{A}=\begin{bmatrix} 1 & 0 \\ 0 & a \end{bmatrix}$, where $a \overset{R}{\leftarrow} Z_p$. 		
		Set $H_1(\mathbf{y})= \hat{g}^{\mathbf{A}r} = \begin{bmatrix} g^{r} & 0 \\ 0 &  g^{ar} \end{bmatrix}\in \hat{G}^{2 \times 2}$, where $r=F''(l)$, and $F''$ is a random function onto $Z_p$.
		It utilizes the multi-DDH assumption which tightly reduces to the DDH assumption using the random-self reducibility.
		We have
		\begin{equation*}
			Adv_1-Adv_2 \leq Adv_{\hat{G}}^{DDH} \left( t+4Q_1 \times t_{\hat{G}} \right),
		\end{equation*}
		where $t_{\hat{G}}$ is the time for performing an exponentiation, and $Q_1$ stands for the number of $RO_1$ query.		
		
		\noindent
		\hangafter=1
		\setlength{\hangindent}{2em}
		\textbf{Game} $\mathcal{G}_3$: this game is same with the $\mathcal{G}_2$, with the exception that the answers of $\textbf{QPKeyGen}$ query are simulated as $SK_{i,\mathbf{y}}
		=( 
		\hat{g_0}^{\mathbf{S}_i \cdot y_i} 	\cdot  \hat{g}^{\mathbf{K}_i \mathbf{A}\mathbf{H}_\mathbf{y}} \cdot \hat{g}^{\mathbf{T}_i \mathbf{H}_\mathbf{y}},
		( \hat{g}^{\mathbf{H}_{\mathbf{y}}} \cdot \hat{g_1}^{B(\gamma)} )^{\mathbf{T}_i},
		g_1^{B(\gamma)\mathbf{T}_i},
		\hat{g}^{\mathbf{H}_{\mathbf{y}}w_i},
		g_2^{\overline{B(\gamma)}\mathbf{T}_i}
		)$ for each $\mathbf{y}$, where $\mathbf{S}_i \overset{R}{\leftarrow} Z_p^{2 \times 2}$ such that $\sum_{i \in B} \mathbf{K}_i = \mathbf{0}$.
		We will illustrate below that 
		\begin{equation*}
			Adv_2 = Adv_3.
		\end{equation*}		
		\noindent  We provide the following new hybrid proof to prove the gap between $\mathcal{G}_2$ and $\mathcal{G}_3$, where index $q_1 \in 1,...,Q_1$.
		
		\noindent
		\hangafter=1
		\setlength{\hangindent}{2em}
		\textbf{Game} $\mathcal{G}_{3.1.1}$: this game is same with the $\mathcal{G}_2$.
		Therefore,
		\begin{equation*}
			Adv_2 = Adv_{3.1.1}.
		\end{equation*}
		
		\noindent
		\hangafter=1
		\setlength{\hangindent}{2em}
		$\mathcal{G}_{3.q_1.1} \leadsto \mathcal{G}_{3.q_1.2}$: 
		$\mathcal{B}_1$ changes the answer of the \textit{q-}th $RO_1$ query from a truly random matrix in the space $\hat{g}^{\mathbf{A}}$ to a random value in $\hat{G}^{2 \times 2}$, applying the DDH assumption.
		$\mathcal{B}_1$ selects randomly $v_1,v_2 \in Z_p$, and utilizes a basis $ \left( \mathbf{A}_1, \mathbf{A}_2 \right) = \left( \begin{bmatrix} a & 0 \\ 0 & 0 \end{bmatrix},  \begin{bmatrix} 0 & 0 \\ 0 & -a \end{bmatrix} \right) $ of $D_2{(Z_p)}$ to express a uniformly random matrix $v_1 \mathbf{A}_1 + v_2 \mathbf{A}_2 $, where $D_2(Z_p)$ denotes all $2$-order diagonal matrices of $Z_p$. 	
		On the $q_1$ query, we change it to $v_1 \mathbf{A}_1 + v_2 \mathbf{A}_2 $, where $v_1 \overset{R}{\leftarrow} Z_p ,v_2 \overset{R}{\leftarrow} Z_p^* $.
		It only changes $\mathcal{A}$'s view by a statistical distance of $1/p$. 
		At the same time, the step with $v_2 \in Z_p^*$ should guarantee $\mathbf{H_y}\mathbf{A}_2 \neq 0$.
		We have
		\begin{equation*}
			Adv_{3.q_1.1} - Adv_{3.q_1.2} \leq Adv_{\hat{G}}^{DDH}(t)+1/p.
		\end{equation*}
		
		\noindent
		\hangafter=1
		\setlength{\hangindent}{2em}
		$\mathcal{G}_{3.q_1.2} \leadsto \mathcal{G}_{3.q_1.3}$: 
		$\mathcal{B}_1$ will simulate the partial functional key $SK_{i,\mathbf{y}}$ by executing the algorithm $\mathbf{PKeyGen}(B,sk_i,\mathbf{y},t)\rightarrow SK_{i,\mathbf{y}}$ with limitation that $|B|=t>|\mathcal{CS}|$.
		There exist three cases:
		\begin{enumerate}
			
			\item For queries \textbf{QPKeyGen} with $\{(i,y_i)\}_{i\in \mathcal{CS}}$, $\mathcal{A}$ generates $\{SK_{i,\mathbf{y}}\}_{i\in \mathcal{CS}}$ by itself.
			
			\item For queries \textbf{QPKeyGen} with $\{(i,y_i)\}_{i\in \mathcal{HS} \setminus B}$, $\mathcal{B}_1$ sets $y_i = 0$.
			
			\item For queries \textbf{QPKeyGen} with $\{(i,y_i)\}_{i\in B\bigcap\mathcal{HS}}$, $\mathcal{B}_1$ will simulate
			$SK_{i,\mathbf{y}}$ as $SK_{i,\mathbf{y}}
			=( 
			\hat{g_0}^{\mathbf{S}_i \cdot y_i} 	\cdot \hat{g}^{F_i(\mathbf{y})} \cdot \hat{g}^{\mathbf{T}_i \mathbf{H_y}},
			\left( \hat{g}^{\mathbf{H}_{\mathbf{y}}} \cdot \hat{g_1}^{B(\gamma)} \right)^{\mathbf{T}_i},
			g_1^{B(\gamma)\mathbf{T}_i},
			\hat{g}^{\mathbf{H}_{\mathbf{y}}w_i},
			g_2^{\overline{B(\gamma)}\mathbf{T}_i}
			)$, where $F_i$ denotes a random function onto $Z_p^{ 2 \times 2 }$ and $\sum_{i \in B} F_i(\mathbf{y})=0$ for any $\mathbf{y}$.
			To achieve this aim, $\mathcal{B}_1$ first prove the two distributions below are identical for any $\mathbf{K}_i \in Z_p^{2 \times 2}$, such that $\sum_{i \in B} \mathbf{K}_i = \mathbf{0}$:
			\begin{equation*}
				\left(\mathbf{T}_i\right)_{i \in B\bigcap\mathcal{HS}} \text{ and } \left( \mathbf{T}_i + \mathbf{K}_i \mathbf{A}_2  \right)_{i \in B\bigcap\mathcal{HS}}.
			\end{equation*}	
			The outputs of $RO_1$ queries:	
			\begin{enumerate}[label={\textbullet}]
				\item On the query $j <$ \textit{q} $RO_1$-query, the answers of $RO_1$ are randomly in the space of $\hat{g}^{\mathbf{A}}$, i.e.,
				$H_2(\mathbf{y})= \hat{g}^{\mathbf{A}r} = \begin{bmatrix} g^{r} & 0 \\ 0 &  g^{ar} \end{bmatrix}\in \hat{G}^{2 \times 2}$.
				
				\item On the \textit{q}-th $RO_1$-query, the answer of $RO_1$ is a random value in $D_2{(\hat{G})}$, i.e.,
				$H_2(\mathbf{y})= F'(\mathbf{y})$.
			\end{enumerate}
			In the $\mathcal{A}$'s view, the extra terms $\left( \mathbf{K}_i \mathbf{A}_2 \right)_{i \in B\bigcap\mathcal{HS}}$ possibly appear as follow:		
			\begin{enumerate}[label={\textbullet}]
				\item On the \textit{q}-th $RO_1$-query, $
				SK_{i,\mathbf{y}}
				=( \hat{g_0}^{\mathbf{S}_i\cdot y_i}\cdot\hat{g}^{\mathbf{T}_i \cdot \mathbf{K}_i \mathbf{A}_2 \mathbf{H_y}}, 
				( \hat{g}^{\mathbf{H_y}} \cdot \hat{g_1}^{B(\gamma)} )^{\mathbf{T}_i \cdot \mathbf{K}_i \mathbf{A}_2}, 
				g_1^{B(\gamma)\mathbf{T}_i \cdot \mathbf{K}_i \mathbf{A}_2 },\\
				\hat{g}^{\mathbf{H}_{\mathbf{y}}w_i},
				g_2^{\overline{B(\gamma)}\mathbf{T}_i \cdot \mathbf{K}_i \mathbf{A}_2}
				)
				$, where $\mathbf{A}_2 \mathbf{H_y} \neq 0$. 
				The matrices $\mathbf{K}_i\mathbf{A}_2 \mathbf{H_y}$ are randomly distributed over $D_2{(Z_p)}$, such that $\sum_{i \in B} \mathbf{K}_i\mathbf{A}_2 \mathbf{H_y} = \mathbf{0}$.
				It is as in $\mathcal{G}_{3.q_1.3}$, by setting $F_i = \mathbf{K}_i\mathbf{A}_2 \mathbf{H_y}$.
				
				\item Otherwise,				
				$
				SK_{i,\mathbf{y}}
				=( \hat{g_0}^{\mathbf{S}_i\cdot y_i}\cdot\hat{g}^{\mathbf{T}_i \cdot \mathbf{K}_i \mathbf{A}_2 \mathbf{H_y}}, 
				( \hat{g}^{\mathbf{H_y}} \cdot \hat{g_1}^{B(\gamma)} )^{\mathbf{T}_i \cdot \mathbf{K}_i \mathbf{A}_2}, 
				g_1^{B(\gamma)\mathbf{T}_i \cdot \mathbf{K}_i \mathbf{A}_2 },
				\hat{g}^{\mathbf{H}_{\mathbf{y}}w_i},
				g_2^{\overline{B(\gamma)}\mathbf{T}_i \cdot \mathbf{K}_i \mathbf{A}_2}
				)
				$.
				Hence, the terms 
				$\hat{g_0}^{\mathbf{S}_i\cdot y_i}\cdot\hat{g}^{\mathbf{T}_i \cdot \mathbf{K}_i \mathbf{A}_2 \mathbf{H_y}}, 
				( \hat{g}^{\mathbf{H_y}} \cdot \hat{g_1}^{B(\gamma)} )^{\mathbf{T}_i \cdot \mathbf{K}_i \mathbf{A}_2},  
				g_1^{B(\gamma)\mathbf{T}_i \cdot \mathbf{K}_i \mathbf{A}_2 },
				g_2^{\overline{B(\gamma)}\mathbf{T}_i \cdot \mathbf{K}_i \mathbf{A}_2}$
				can provide $\left( \mathbf{K}_i \mathbf{A}_2 \right)_{i \in B\bigcap\mathcal{HS}}$ for $\mathcal{A}$.

			\end{enumerate}


		\end{enumerate}
		
		\noindent
		\hangafter=1
		\setlength{\hangindent}{2em}
		$\mathcal{G}_{3.q_1.3} \leadsto \mathcal{G}_{3.q_1+1.1}$: 
		We utilize the DDH instance to replace the distribution of the $\hat{g}^{\mathbf{H}_{\mathbf{y}}}$ to uniformly random in the space of $\hat{g}^{\mathbf{A}}$.
		From the $v_1 \mathbf{A}_1 + v_2 \mathbf{A}_2  $ to a random matrix in the space of $\hat{g}^{\mathbf{A}}$ only changes $\mathcal{A}$'s view by a statistical distance of $1/p$, which is the reverse of $\mathcal{G}_{3.q_1.1} \leadsto \mathcal{G}_{3.q_1.2}$. 
		Therefore,  
		\begin{equation*}
			Adv_{3.q_1.3} - Adv_{3.q_1+1.1} \leq Adv_{\hat{G}}^{DDH}(t)+1/p.
		\end{equation*}			
		We notice that $\mathcal{G}_{3.Q_1+1.1} = \mathcal{G}_{4.0}$.	
		All the $SK_{i,\mathbf{y}}$ output by \textbf{QPKeyGen} can be calculated only obtaining $\sum_{i \in B} \mathbf{S}_i y_i$.
		Hence, $B(\gamma)\mathbf{T}_i$ and $F_i(\mathbf{y})$ mask the vectors $\mathbf{y}_iy_i$.
		For correct decryption,  $\sum_{i \in B}\mathbf{S}_i y_i$ must be revealed and the equation $B(\gamma)\cdot \gamma^t = \overline{B(x)}$ holds.
		$\mathcal{B}_1$  selects $\mathbf{T}_i \in D_2{(Z_p)}$ randomly. 
		$\mathcal{B}_1$  answers oracle calls to $RO_1$ and \textbf{QPKeyGen} from $\mathcal{A}$ utilizing its own oracle.	
		\begin{enumerate}
			
			\item if $i$ is the last honest index in $B$, $\mathcal{B}$ responses   $SK_{i,\mathbf{y}}
			=( 
			\hat{g_0}^{\mathbf{S}_i \cdot y_i} 	\cdot \hat{g}^{F_i(\mathbf{y})} \cdot \hat{g}^{\mathbf{T}_i \mathbf{H_y}},
			( \hat{g}^{\mathbf{H}_{\mathbf{y}}} \cdot \hat{g_1}^{B(\gamma)} )^{\mathbf{T}_i},
			g_1^{B(\gamma)\mathbf{T}_i},
			\hat{g}^{\mathbf{H}_{\mathbf{y}}w_i},
			g_2^{\overline{B(\gamma)}\mathbf{T}_i}
			)$
			to $\mathcal{A}$.
			
			\item otherwise, $\mathcal{B}$ selects a polynomials $B'(x)$ with degrees less than $t$, and responses $SK_{i,\mathbf{y}}
			=( 
			\hat{g}^{F_i(\mathbf{y})} \cdot \hat{g}^{\mathbf{T}_i \mathbf{H_y}},
			( \hat{g}^{\mathbf{H}_{\mathbf{y}}} \cdot \hat{g_1}^{B'(\gamma)} )^{\mathbf{T}_i},
			g_1^{B'(\gamma)\mathbf{T}_i},
			\hat{g}^{\mathbf{H}_{\mathbf{y}}w_i},
			g_2^{\overline{B'(\gamma)}\mathbf{T}_i}
			)$
			to $\mathcal{A}$.
		\end{enumerate}	
		
		\noindent
		\hangafter=1
		\setlength{\hangindent}{2em}
		\textbf{Game} $\mathcal{G}_4$:  We change the distribution of the matrices $g^{\mathbf{H_{\mathbf{y}}}}$, which are generated by $RO_1$, back from being uniformly random within the space of $g^{\mathbf{A}}$ to being uniformly random over $\hat{G}^{2 \times 2}$, thereby returning to $H_1(\mathbf{y})$.
		We have 
		\begin{equation*}
			Adv_3-Adv_4 \leq Adv_{\hat{G}}^{DDH} \left( t+4Q_1 \times t_{\hat{G}} \right).
		\end{equation*}	
		\noindent We provide the following games for simulating $RO_2$.
		
		\noindent
		\hangafter=1
		\setlength{\hangindent}{2em}
		\textbf{Game} $ \mathcal{G}_{5} $:
		We introduce a simulator $\mathcal{B}_2$ who simulates output of $RO_2$ through a process similar with $RO_1$.
		It simulates any new output of $RO_2$ as a truly random matrix in the space $g^{\mathbf{D}}$ for $\mathbf{D}=\begin{bmatrix} 1 & 0 \\ 0 & d \end{bmatrix}$, where $d \overset{R}{\leftarrow} Z_p$ and utilizes the multi-DDH assumption.
		Hence, we have
		\begin{equation*}
			Adv_{4}-Adv_{5} \leq Adv_{G}^{DDH} \left( t+4Q_2 \times t_{G} \right),
		\end{equation*}
		where $t_G$ is the time for performing an exponentiation, and $Q_2$ denotes the number of $RO_2$ query.
		
		\noindent
		\hangafter=1
		\setlength{\hangindent}{2em}
		\textbf{Game} $ \mathcal{G}_{6}$:
		$\mathcal{B}_2$ simulates any \textbf{QEncrypt} with $x_i^{0}$ instead of $x_i^b$, 
		In addition, $\mathcal{B}_2$ change the answers of any $RO_2$ query to a random matrix in $G^2$.	
		To achieve this aim, we introduce a hybrid game as below, where index $q_2 \in 1,...,Q_2$. 	
		
		\noindent
		\hangafter=1
		\setlength{\hangindent}{2em}
		$\mathcal{G}_{6.1,1}$: is same with $\mathcal{G}_{5}$.
		We have 
		\begin{equation*}
			Adv_{5} = Adv_{6.1,1}.
		\end{equation*}
		
		\noindent
		\hangafter=1
		\setlength{\hangindent}{2em}
		$\mathcal{G}_{6.q_2.1} \leadsto \mathcal{G}_{6.q_2.2}$: is similar with $\mathcal{G}_{3.q_1.1} \leadsto \mathcal{G}_{3.q_1.2}$. 
		It utilizes the DDH assumption to change the distribution of answers of the \textit{q}-th $RO_2$ from the space of $g^{\mathbf{D}}$ to a random matrix over $G^{2 \times 2}$.
		The basis $(\mathbf{D}_1,\mathbf{D}_2) = \left( \begin{bmatrix} d & 0 \\ 0 & 0 \end{bmatrix},  \begin{bmatrix} 0 & 0 \\ 0 & -d \end{bmatrix} \right)$  $ \in D_2{(Z_p)}$ is used to build a random matrix $u_1 \cdot \mathbf{D}_1+u_2 \cdot \mathbf{D}_2$ where $u_1,u_2 \overset{R}{\leftarrow} Z_p$.
		Notice that $\mathbf{H}_l \mathbf{D}_2 \neq 0$.
		On the $q_2$ query, we change it to $u_1 \cdot \mathbf{D}_1+u_2 \cdot \mathbf{D}_2$, where $u_1 \overset{R}{\leftarrow} Z_p,u_2 \overset{R}{\leftarrow} Z_p^*$.
		We have 
		\begin{equation*}
			Adv_{6.q_2.1} - Adv_{6.q_2.2} \leq Adv_{G}^{DDH}(t)+1/p.
		\end{equation*}
		
		\noindent
		\hangafter=1
		\setlength{\hangindent}{2em}
		$\mathcal{G}_{6.q_2.2} \leadsto \mathcal{G}_{6.q_2.3}$: $\mathcal{B}_2$ will change the ciphertexts 
		$CT_{i,l,t}=( 
		(g^{\mathbf{H}_l})^{\mathbf{S}_i} \cdot g_0^{x_i^b}, 
		( g^{\mathbf{H}_l}\cdot g_1^{w_i} )^{\mathbf{T}_i}, 
		( g_1\cdot g_2^{\gamma^{t}} )^{w_i}, 
		\hat{g_1}^{w_i},
		g_2^{\gamma^{t}\cdot \mathbf{T}_i}
		)$ 
		as 
		$CT_{i,l,t}=( 
		(g^{\mathbf{H}_l})^{\mathbf{S}_i} \cdot g_0^{x_i^0}, 
		( g^{\mathbf{H}_l}\cdot g_1^{w_i} )^{\mathbf{T}_i}, 
		( g_1\cdot g_2^{\gamma^{t}} )^{w_i}, 
		\hat{g_1}^{w_i},
		g_2^{\gamma^{t}\cdot \mathbf{T}_i}
		)$.	
		We will prove this does not change the view of $\mathcal{A}$.
		\begin{enumerate}
			\item $\mathcal{B}_2$ first guesses $\mathbf{z}_i \overset{R}{\leftarrow} Z_p^2$ for all $i \in [n]$.
			Each $\mathbf{z}_i$ is either a pair $\left( x_i^0,x_i^1 \right)$ resulting from a query to \textbf{QEncrypt}, or $\perp$ that stands for no query to \textbf{QEncrypt}.
			If $\mathcal{B}_2$ guesses successfully, it simulates the view of $\mathcal{A}$ actually. We call this case $E$ event.
			Otherwise, the guess was unsuccessful, the simulation aborts, and $\mathcal{B}_2$ outputs a value $\alpha \in \{0,1\}$ randomly.		
			The event $E$ is independent of the view of $\mathcal{A}$, and happens with probability $(p^2+1)^{-1}$.
			
			\item If $\left( \mathbf{z}_i \right)_{i \in [n]}$ are consistent, the game goes on normally. 
			Guess $b'$ on $b$ will be outputted ($E'$ event).
			On $E'$ event, $\mathcal{G}_{6.q_2.2} $ and $ \mathcal{G}_{6.q_2.3}$ are distributed identically:
			given two tuples 
			$
				(\mathbf{S}_i)_{i \in [B], \mathbf{z}_i = (x_i^0, x_i^1)} $ and $ \left(\mathbf{S}_i + \mathbf{D}_2 \cdot \gamma(x_i^b - x_i^0)\right)_{i \in [B], \mathbf{z}_i = (x_i^0, x_i^1)},
			$
			where $\gamma \overset{R}{\leftarrow} Z_p$, and $\mathbf{S}_i \overset{R}{\leftarrow} D_2{(Z_p)}$ for all $i$.
			We show re-write $\mathbf{S}_i$ as $\mathbf{S}_i + \mathbf{D}_2 \cdot \gamma(x_i^b - x_i^0)$ while maintaining the same distribution in this game.
			\begin{enumerate}
				\item Extra term $\mathbf{D}_2 \cdot \gamma(x_i^b - x_i^0)$ might be found in \textbf{QPKeyGen} as 
				$
					SK_{i,\mathbf{y}}
					=( \hat{g_0}^{\mathbf{S}_i\cdot y_i + \mathbf{D}_2 \cdot \gamma(x_i^b - x_i^0) }\cdot\hat{g}^{\mathbf{T}_i \mathbf{h}_\mathbf{y}},
					( \hat{g}^{\mathbf{H}_{\mathbf{y}}} \cdot \hat{g_1}^{B(\gamma)} )^{\mathbf{T}_i},
					g_1^{B(\gamma)\mathbf{T}_i},
					\hat{g}^{\mathbf{H}_{\mathbf{y}}w_i},
					g_2^{\overline{B(\gamma)}\mathbf{T}_i}
					).
				$
				The $\sum_{i \in B} \mathbf{D}_2 \cdot \gamma(x_i^b - x_i^0)y_i = \mathbf{0}$, because  $f(\mathbf{x}^0)=f(\mathbf{x}^1)$.
				
				\item Extra term $\mathbf{D}_2 \cdot \gamma(x_i^b - x_i^0)$ might appear in the output of \textbf{QEnc} on $q-$th $RO_2$ query,  because the $H_2(l)$ lies in the space of $g^\mathbf{D}$.
				We have 
				$
					CT_{i,l,t}=( 
					\left(g^{\mathbf{H}_l}\right)^{\mathbf{S}_i + \mathbf{D} \cdot \gamma(x_i^b - x_i^0)} \cdot g_0^{x_i^b}, 
					( g^{\mathbf{H}_l}\cdot g_1^{w_i} )^{\mathbf{T}_i}, 
					( g_1\cdot g_2^{\gamma^{t}} )^{w_i}, 
					\hat{g_1}^{w_i},
					g_2^{\gamma^{t}\cdot \mathbf{T}_i}
					).
				$
				It sets $\gamma = -\frac{1}{\mathbf{H_l} \mathbf{D}}$ mod $p$, and hence,
				$
					CT_{i,l,t}=( 
					(g^{\mathbf{H}_l})^{\mathbf{S}_i} \cdot g_0^{x_i^0}, 
					( g^{\mathbf{H}_l^{\top}}\cdot g_1^{w_i} )^{\mathbf{T}_i}, 
					\left( g_1\cdot g_2^{\gamma^{t}} \right)^{w_i}, 
					\hat{g_1}^{w_i},
					g_2^{\gamma^{t}\cdot \mathbf{T}_i}
					)
				$, which is the encryption of $x_i^0$.
				$\gamma$ is  independent of the $i$, hence, it  converts all the encryptions of $x_i^b$ into $x_i^0$.
				
			\end{enumerate}
			
			Hence, the games are identically when $E'$ happens. 
			Otherwise, the games outputs $\alpha \in \{0,1\}$.
			We have 
			\begin{equation*}
				Adv_{6.q_2.2} = Adv_{6.q_2.3}.
			\end{equation*}
			
			\hangafter=1
			\setlength{\hangindent}{2em}
			$\mathcal{G}_{6.q_2.3} \leadsto \mathcal{G}_{6.q_2+1.1}$:
			This transition is similarly reverse $\mathcal{G}_{6.q_2.1} \leadsto \mathcal{G}_{6.q_2.2}$.
			We switch back the distribution of $g^{\mathbf{H}_l}$ on the $q_2$-th $RO_2$-query from a uniform random distribution  over $D_2{(G)}$ to a uniform random distribution in the space of $g^{\mathbf{D}}$ by using the DDH assumption.
			We have 
			\begin{equation*}
				Adv_{6.q_2.3} - Adv_{6.q_2+1.1} \leq Adv_{G}^{DDH}(t)+1/p.
			\end{equation*}
			
			$\mathcal{G}_{6.q_2+1.1} = \mathcal{G}_{7}$, and $Adv_{7} = 0$, we have
			\begin{equation*}
				Adv_{6} - Adv_{7} \leq 2 Q_2 (Adv_{G}^{DDH}(t)+1/p).
			\end{equation*}	
		\end{enumerate}		
	\end{IEEEproof}

\section{Experimental Analysis} \label{Experimental}

	\begin{figure*}[!t]
		\centering
		
		\subfloat[$\mathbf{Setup}$]{
			\includegraphics[width=3.0in]{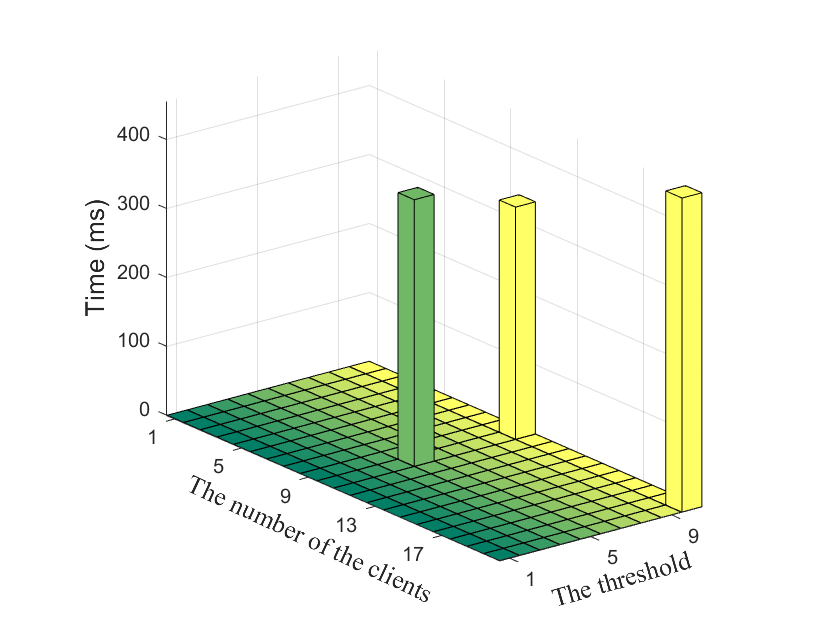}
			\label{setupTime}
		}
		\hfil
		\subfloat[$\mathbf{PKeyGen}$]{
			\includegraphics[width=3.0in]{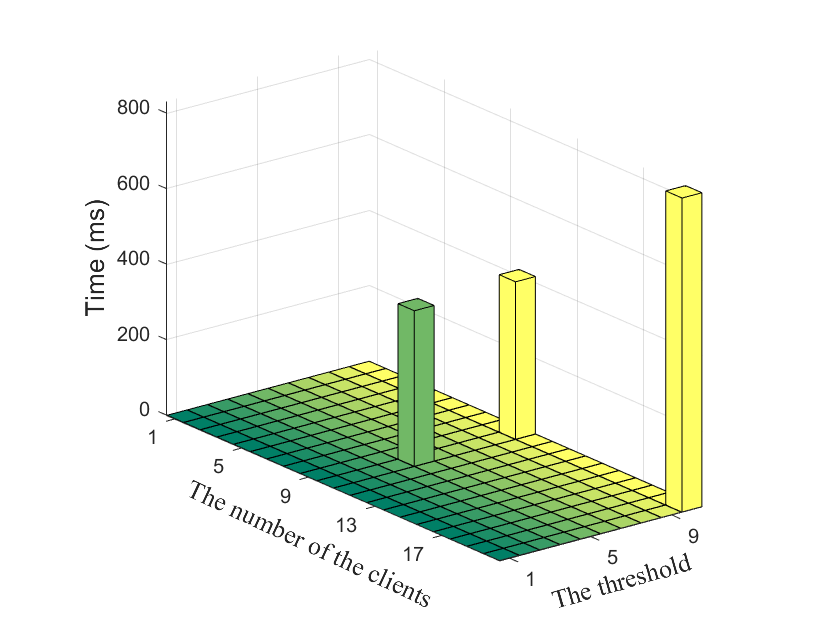}
			\label{keygenTime}
		}		
		\hfil
		\subfloat[$\mathbf{Enc}$]{
			\includegraphics[width=3.0in]{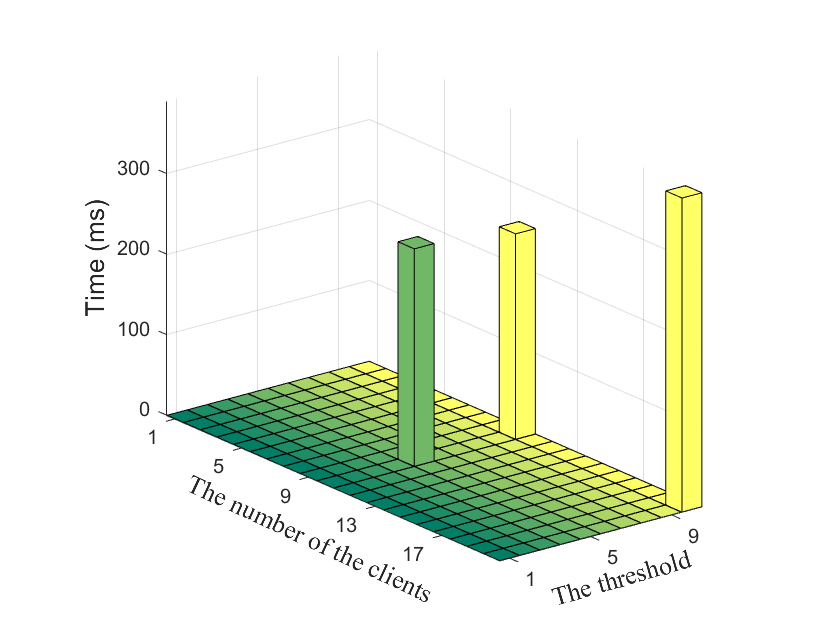}
			\label{encTime}
		}
		\hfil
		\subfloat[$\mathbf{Dec}$]{
			\includegraphics[width=3.0in]{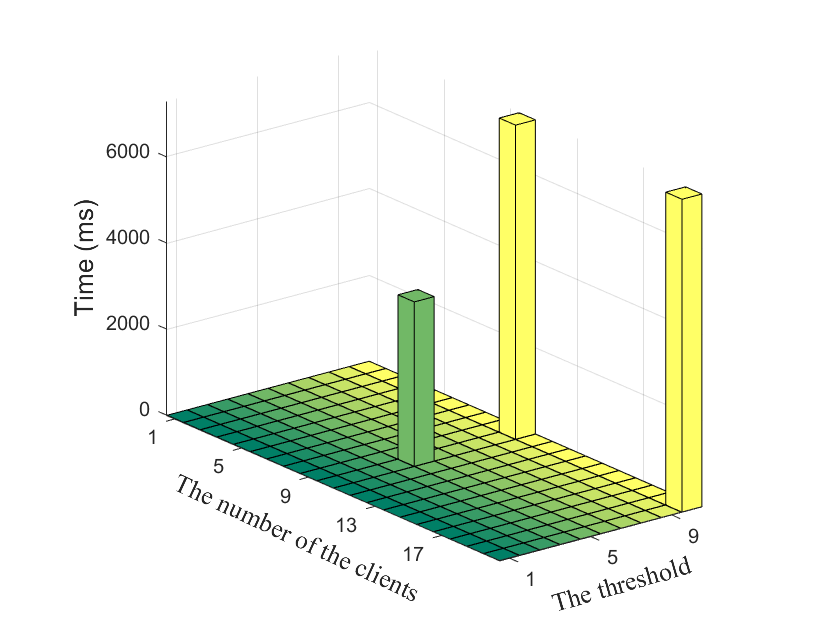}
			\label{decTime}
		}
		\caption{The computational cost of each algorithm in the FTMCFE-IP scheme.}
		\label{time}		
	\end{figure*}
	
	We implement our FTMCFE-IP scheme on a Lenovo Y9000K laptop equipped with an Intel i7-11800H CPU and 32G RAM.
	We apply the JPBC library \cite{JPBC}  for implementing the bilinear map.
	This library is open-source, implemented in Java, and supports various elliptic curves as well as other types of algebraic curves.	
	The F curve $y^2 = x^3 + b$ is a pairing-friendly elliptic curve of prime order, and can support Type-III pairing.
	We select the F curve to implement each algorithm of our FTMCFE-IP scheme, and choose SHA-256 as the hash function $H_1$ and $H_2$.
	The computation costs of all algorithm are shown in the Fig. \ref{time}.
	
	Let $n$ denote the number of clients, and $t$ represent the threshold value (i.e., the size of online clients $B$ is $t$).
	We consider the following three cases:
	Case-i. $n = 10$, $t = 5$. Case-ii. $n = 10$, $t = 10$. Case-iii. $n = 20$, $t = 10$.
	For every case, every algorithm is executed ten times, and the average result is reported as the experimental result.
	
	In the $\mathbf{Setup}$ algorithm, TA calculates public parameters $\{g_1^{\gamma^{j}}\}_{j\in[n]}, \{g_2^{\gamma^{j}}\}_{j\in[n]}, \{\hat{g_1}^{\gamma^{j}}\}_{j\in[n]}$, and  each $CL_i$ generates separately its secret key $sk_i$ and encryption key $ek_i$, where $i \in [1,n]$.	 
	The $\mathbf{Setup}$ algorithm requires approximately 387 ms in Case-i, 337 ms in Case-ii, and 456 ms in Case-iii.
	The computational cost of $\mathbf{Setup}$ algorithm is linear with  $n$.
	
	In the $\mathbf{PKeyGen}$ algorithm, each client generates the partial functional keys for the aggregator. 
	The computational costs of $\mathbf{PKeyGen}$ algorithm are approximately 412 ms in Case-i, 417 ms in Case-ii, and 832 ms in Case-iii.  
	The computation costs of $\mathbf{PKeyGen}$ algorithm grow linearly with $n$.
	
	In the $\mathbf{Enc}$ algorithm, each client applies its encryption key $ek_i$ and a threshold value $t$ to encrypt data independently.
	It takes approximately 270 ms, 255 ms and 390 ms in Case-i, Case-ii and Case-iii, respectively.
	The computational cost of $\mathbf{Enc}$ algorithm increases linearly with $n$.	
	
	The $\mathbf{Dec}$ algorithm is performed by the aggregator.
	When the number of received partial keys reaches the specified threshold $t$, the aggregator can learn an inner product.
	$\mathbf{Dec}$ algorithm takes approximately 3819 ms, 7297 ms and 7628 ms in the three cases above, respectively, which is linear with the threshold $t$.	
	
\begin{figure*}[!t]
		\centering
		\includegraphics[width=0.9\textwidth]{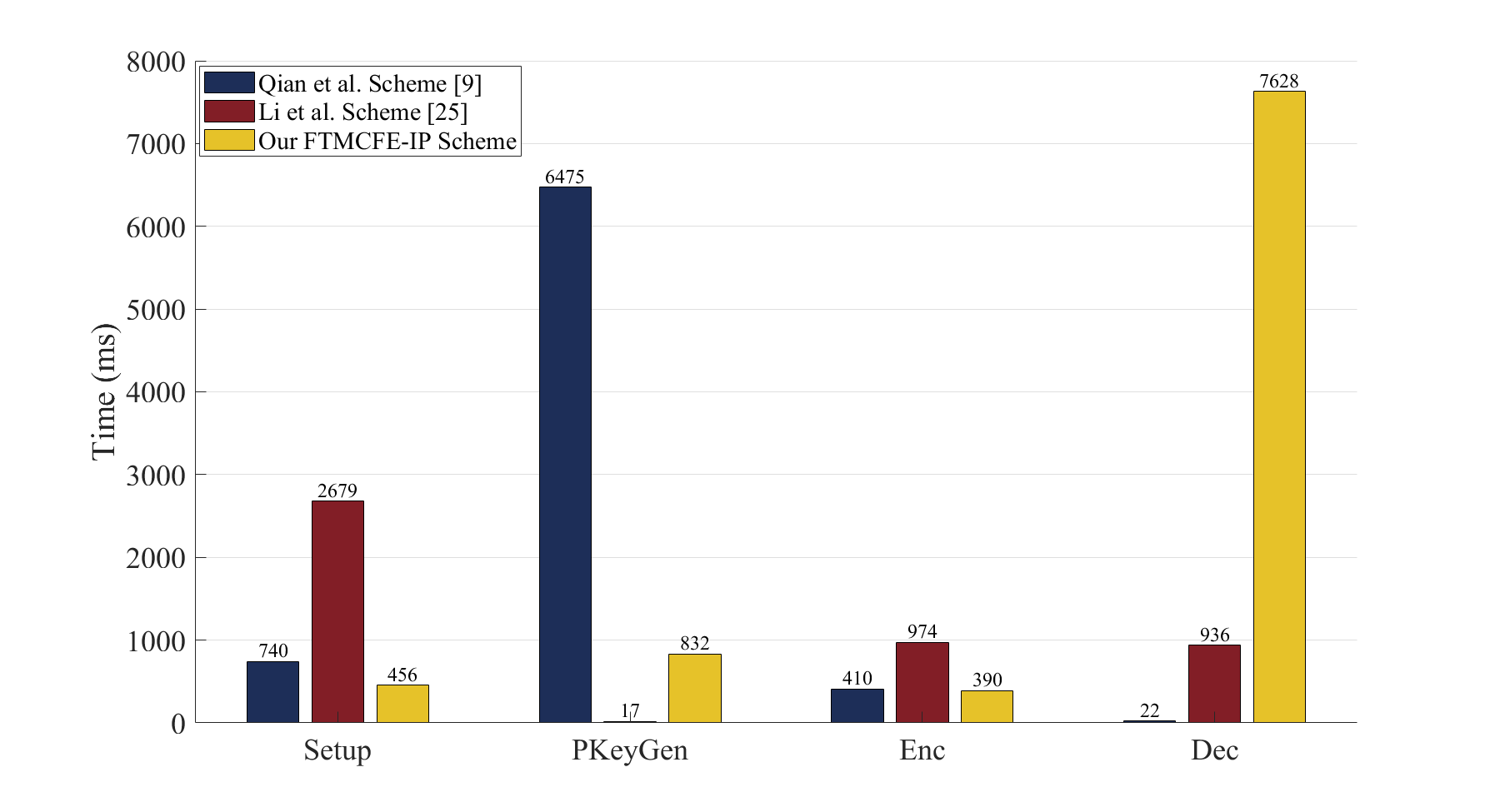} 
		\caption{Efficiency comparison}
		\label{comparison}
	\end{figure*}

	We compare the state-of-the-art schemes \cite{QianDMCFE2024} \cite{LiRobust2023} with our FTMCFE-IP scheme, and implement all algorithms under $n=20$ and $t=10$.	
	The schemes \cite{QianDMCFE2024} \cite{LiRobust2023} are implemented with the following settings.
	\begin{itemize}[label=-]	
		\item Qian et al. scheme \cite{QianDMCFE2024}: 
		to achieve equivalent secure level, we select the 3072-bit integer $N$ for the scheme \cite{QianDMCFE2024}, which is constructed based on the decision composite residuosity  group.
		For realizing cryptographic primitives used in \cite{QianDMCFE2024}, we select the short signature \cite{ShortS2004} as the digital signature scheme, and select AES-128 \cite{AES1999} as the symmetric encryption scheme, and use the Diffie–Hellman key agreement scheme \cite{DH}, and select Shamir's scheme \cite{Shamir1979} as the secret sharing scheme.
		
		\item Li et al. scheme \cite{LiRobust2023}: 
		the cryptographic primitives used in scheme \cite{LiRobust2023} include a special MCFE scheme, a non-interactive key exchange scheme and a secret sharing scheme.
		For realize these primitives, we choose the MCFE scheme from \cite{DMCFE2018}, and select the non-interactive key exchange scheme based on identity-based encryption \cite{NIKE2018}, and alse select Shamir's scheme \cite{Shamir1979} as the secret sharing scheme.
	\end{itemize}		
	The comparison results are shown in the Fig \ref{comparison}.	
	
	The $\mathbf{Setup}$ algorithms in \cite{QianDMCFE2024} and \cite{LiRobust2023} require 740 ms and 2679 ms, respectively, while the $\mathbf{Setup}$ algorithm in our scheme takes 456ms. 	
	In \cite{QianDMCFE2024}, $\mathbf{Setup}$ phase involves the initialization of multiple cryptographic primitives,
	and the scheme \cite{LiRobust2023} additionally requires each client to execute the key agreement to generate sharing keys, which results in more computation overhead.
	Therefore, the $\mathbf{Setup}$ algorithm in our scheme is more efficient.	 	
	In the $\mathbf{PKeyGen}$ phase, the scheme \cite{QianDMCFE2024} requires multiple clients to perform key generation and agreement of the key agreement protocol, execute secret sharing, and individually sign the generated shares with the digital signature, which costs 6475 ms.
	In contrast, the scheme \cite{LiRobust2023} achieves higher efficiency in $\mathbf{PKeyGen}$ phase, which only costs 17 ms.	
	However, the $\mathbf{Enc}$ algorithm in \cite{LiRobust2023} is relatively inefficient, which requires 974 ms and is twice the running time of our scheme.
	The $\mathbf{Enc}$ algorithm in our scheme only costs 390 ms, and  can support each client to encrypt its local data efficiently.	
	A limitation of our FTMCFE-IP scheme is that the $\mathbf{Dec}$ algorithm involves many pairing operations, thereby costing 7628 ms and incurring a comparatively high computational overhead.	
	The $\mathbf{Dec}$ algorithm in the scheme \cite{QianDMCFE2024} and \cite{LiRobust2023} cost 22 ms and 936 ms, respectively.
	
\section{Conclusion} \label{Conclusion}
	We designed a FTMCFE-IP scheme in this paper, which realizes flexible threshold setting and client dropout.
	Our FTMCFE-IP scheme allows multiple clients co-exist and generate ciphertexts locally, which is suitable for FL frameworks.
	The designed scheme supports an aggregator to aggregate clients' ciphertexts, and only obtain a value of inner product while revealing nothing else about raw data.
	In addition, our FTMCFE-IP scheme does not require all clients to be online.
	The aggregator can decrypt successfully if and only if the number of keys which he/she obtained from clients is no less than the threshold.
	The clients are allowed to drop out, which solves the problem of equipment damage or offline in practical applications.
	We provided the formal definition of our FTMCFE-IP scheme, and formalized its security model.
	Moreover, we built the detailed construction, and presented the security proof.	
	Finally, performance analysis shows our scheme is effective.
	
	Our future work is to design function-hiding FTMCFE-IP schemes which prevent functions from being leaked, and to build a privacy-enhanced FL framework.

\section*{Acknowledgments}
	This work was supported by the National Natural Science Foundation of China (Grant No. 62372103), the Natural Science Foundation of Jiangsu Province (Grant No. BK20231149) and the Jiangsu Provincial Scientific Research Center of Applied Mathematics (Grant No.BK202330020).

\bibliography{bare_jrnl_new_sample4}
\bibliographystyle{IEEEtran}

%
%
%
%
%
%

\end{document}